\documentclass[a4paper,10pt]{article}
\usepackage{mathrsfs}
\usepackage{epsfig}
\usepackage{amsmath}
\usepackage{amssymb}

\newcommand{\bea}{\begin{eqnarray}}
\newcommand{\eea}{\end{eqnarray}}
\newcommand{\bnn}{\begin{eqnarray*}}
\newcommand{\enn}{\end{eqnarray*}}
\newcommand{\be}{\begin{equation}}
\newcommand{\ee}{\end{equation}}

\def\PACS{\par\leavevmode\hbox {\it PACS:\ }}%
\def\MSC{\par\leavevmode\hbox {\it MSC:\ }}%
\def\UK{\par\leavevmode\hbox {\it Keywords:\ }}%
\input xy
\xyoption {all}

\begin{document}

\title{Considerations on the hyperbolic complex Klein-Gordon equation}

\author{S. Ulrych\\ Wehrenbachhalde 35, CH-8053 Z\"urich, Switzerland}
%\author{S. Ulrych}\affiliation{Wehrenbachhalde 35, CH-8053 Z\"urich, Switzerland}
\date{May 16, 2010}
\maketitle

\begin{abstract}
The article summarizes and consolidates investigations on hyperbolic
complex numbers with respect to the Klein-Gordon
equation for fermions and bosons. The hyperbolic complex numbers are applied in the
sense that complex extensions of groups and algebras are performed not
with the complex unit, but with the product of complex and hyperbolic
unit. The modified complexification is the key ingredient for the
theory. The Klein-Gordon equation is represented in this framework in the form of the
first invariant of the Poincar\'e group, the mass operator, in order to
emphasize its geometric origin. The possibility of new interactions arising from 
hyperbolic complex gauge transformations is discussed.

\end{abstract}
%\pacs{03.65.Pm; 11.30.Cp; 02.10.Hh; 11.30.Ly; 95.30.Sf}
%\keywords{Hyperbolic complex numbers; Split-complex numbers; Wave equations; Poincar\'e invariance; Hyperbolic Pauli algebra}

%\maketitle
{\scriptsize\PACS{03.65.Pm; 11.30.Cp; 02.10.Hh; 11.30.Ly; 95.30.Sf}
\MSC{81R20; 81R05; 11E88; 15A33; 83D05}
\UK{Hyperbolic complex numbers; Split-complex numbers; Wave equations; Poincar\'e invariance; Hyperbolic Pauli algebra}}

\section{Introduction}
The interest in quaternions and their application to quantum physics 
has been growing considerably since the work of Finkelstein et al. \cite{Fin62,Fin63a,Fin63b}
and the publication of the textbook
of Adler \cite{Adl95} some years ago. Though quaternions could never
compete with the success of vector and matrix approaches, it is hoped
that quaternions might provide deeper insight into the fundamental structures
of physics compared to the conventional representations.

In relativistic quantum physics a quaternionic approach to the
concepts of Dirac has been proposed by Lanczos \cite{Lan26,Lan29a,Lan29b,Lan29c}
in the late twenties of the last century. 
The Dirac equation on the complexified field has been
considered by Conway \cite{Con37}, 
more recent representations have been given by Edmonds \cite{Edm72,Edm84} and 
Gough \cite{Gou86,Gou87,Gou89}. 
Investigations in this area have been performed
also by G\"ursey \cite{Gur50}, who interpreted the 
doubling of the solutions, which is characteristic for quaternionic
representations of the Dirac equation, as isospin \cite{Gur58}.
Other representations of the Dirac equation have been given by
Rotelli \cite{Rot89} and De Leo \cite{Leo98}.
More references about this research area can be found in the work of
Gsponer and Hurni \cite{Gsp01,Gsp08}.

The hyperbolic complex numbers, also called split-complex or double numbers, have become
popular over the past years, though they have been invented already
in the 19th century. They can be seen as a counterpart to the
complex numbers in the sense that they provide relationships
to represent the hyperbolic complex exponential with the help of the hyperbolic unit as a sum of hyperbolic sine and hyperbolic cosine,
analogous to the relationship between sine, cosine, and the complex exponential function \cite{Yag79}.
The notion of hyperbolic or hyperbolic complex numbers is favored by the author in the context of
the hyperbolic functions mentioned above.
The notion of split-complex or double numbers seems to be suitable for the null-plane
representation of the hyperbolic complex numbers \cite{Huc93}.

A recent overview of the existing hyperbolic mathematics with
physical flavour is given by Catoni et al. \cite{Cat08}.
A brief review on hyperbolic complex analysis has been published earlier by
Lambert et al. \cite{Lam87, Lam88}.
The hyperbolic complex numbers have been applied to the representation of the
Schr\"odinger and Klein-Gordon equation by Bracken and Hayes \cite{Bra03}.
Hyperbolic Fourier analysis and hyperbolic interferences have been investigated by Khrennikov and co-worker \cite{Khr07,Khr06}.
For a generalization of the concepts of complex and hyperbolic complex number plane
see the work of Kisil \cite{Kis07}. 

These publications consider the hyperbolic complex numbers mainly 
in the context of hyperbolic complex analysis.
In addition, one can use the hyperbolic unit also to introduce a new 
representation of the real Clifford algebra $\mathbb{R}_{3,0}$ \cite{Ulr05Fund}.
The Clifford algebra $\mathbb{R}_{3,0}$ is a natural framework for 
relativistic physics, e.g., the theory of electrodynamics has been transformed by Baylis \cite{Bay99,Bay04}
into this framework using the Pauli algebra as the explicit representation for $\mathbb{R}_{3,0}$. 
The isomorphic hyperbolic complex representation of $\mathbb{R}_{3,0}$
makes it possible to define conjugation in a form that agrees with
common conventions in physics \cite{Ulr08}, and thus allows one to embed this
algebra smoothly into the existing framework of theoretical physics. 

A Clifford algebra is concerned with rotations and spin.
The hyperbolic unit is therefore used in Ref.~\cite{Ulr05Fund}
to describe the spin structure of a physical state,
whereas the dynamics as part of an infinite Hilbert space, including momentum or orbital angular momentum, 
is covered by conventional complex analysis.
Attempts to generalize the orbital angular momentum to hyperbolic complex
analysis as proposed in Ref.~\cite{Ulr05Fund} seem to have no relevance,
as these generalizations are not solutions of the Klein-Gordon equation
in the corresponding parameterization.

The $\mathbb{R}_{3,0}$ algebra in the hyperbolic complex representation
can be understood as the complexification 
of the quaternions not with the complex unit, 
but with the product of complex and hyperbolic unit, thus providing the relation to the quaternionic models mentioned in the beginning of this section.
The hyperbolic complex extension, which appears also in the context of the group structure,
is actually the key to the mathematical framework proposed in Ref.~\cite{Ulr05Fund}.

It is a natural question, whether these hyperbolic complex extensions of the
group structure also appear in the context of gauge groups. It has been proposed
in Ref.~\cite{Ulr06Gravi} to understand a Maxwell-like theory of gravitation as the hyperbolic complex extension
of electromagnetism. Consequently, there should also be hyperbolic complex counterparts 
for the other interactions. However, there are currently no obvious experimental indications
for this assumption. Nevertheless, even if these hyperbolic
complex extensions have no relevance for the internal symmetries
of a particle state, it should be the goal of
future work to understand exactly why.

\section{Hyperbolic Pauli Algebra}
\label{alg}
The theory is based on a relativistic paravector algebra, denoted in the following as $W$, 
which is used to represent relativistic space-time. A relativistic
vector $x\in W$ is expanded as
\be
´x=x^\mu e_\mu,\hspace{0.5cm}x^\mu\in\mathbb{R}^{\,3,1}\;.
\ee
The space-time metric is realized by the 
basis elements, which are defined as 
\be
\label{basis}
e_\mu=(e_0,e_k)=(1, j\sigma_k)\;. 
\ee
The paravector algebra is formed by the Pauli algebra $\sigma_k$. However, each basis element of the Pauli algebra is multiplied
by the hyperbolic unit $j\equiv \sqrt{+1}$.
The hyperbolic Pauli algebra $e_k$ is thus still isomorphic to the Pauli algebra $\sigma_k$, but
includes the hyperbolic unit as an extra factor. The multiplication rules of the hyperbolic Pauli algebra are
\be
e_ke_l=ij\,\varepsilon_{klm}e^m\;.
\ee
In the same way as the Pauli algebra can be understood as the complexification
of the quaternions by the complex unit $i$,
the hyperbolic Pauli algebra can be understood as a complexification of
the quaternions by the factor $ij\equiv\sqrt{-1}\sqrt{+1}$.

Rotations and boosts of an element $x\in W$ are
realized with the help of the reversion anti-involution
\be
\label{trafo}
x\rightarrow x^\prime=gxg^\dagger\;,
\ee
where $g$ is an element of the spin group. The details of the spin group will be discussed in Sec.~\ref{spaces}.
Reversion leaves the sign of the basis elements unchanged 
\be
e_k^\dagger=e_k
\ee
 and reverses
the order of the elements 
\be
\label{reverse}
(e_ke_l)^\dagger=e_le_k\;.
\ee
One may introduce another anti-involution, called conjugation, which changes the sign of the
basis elements
\be
\bar{e}_k= -e_k
\ee
and reverses the order of the elements as in Eq.~(\ref{reverse}).

A scalar product can be introduced with the trace operator and the
explicit matrix representation of the relativistic paravector algebra,
which is obtained by representing the Pauli algebra in Eq.~(\ref{basis}) with the
Pauli matrices
\be
(x,y)\rightarrowtail \frac{1}{2}tr(\bar{x}y)\in \mathbb{R}\;.
\ee

In the matrix representation conjugation corresponds to a change in sign of 
complex and hyperbolic unit and transposition of the matrix, whereas 
reversion changes only the sign of the complex unit plus transposition
of the matrix. Reversion thus corresponds to the usual Hermitian conjugation.
It is one of the main benefits of the hyperbolic Pauli algebra that the difference between
reversion and conjugation can be represented in a simple and clear form.
Furthermore, conjugation is now defined in accordance
with common conventions in physics in terms of matrix transposition
and reversion of sign of the involved complex and hyperbolic units,
which is not the case in the conventional representation of the Pauli algebra \cite{Por95}.

The terminology of reversion and conjugation refers to the Clifford algebra approach.
The hyperbolic Pauli algebra corresponds to the real Clifford algebra
$\mathbb{R}_{3,0}$. The representation of $\mathbb{R}_{3,0}$ in terms of the Pauli 
algebra has been applied by Baylis to electrodynamics \cite{Bay99}.
Its usefulness in concrete physical applications therefore does not need
to be justified further. The results from Baylis can be adopted on a one to one basis by
the hyperbolic Pauli algebra approach.

In the hyperbolic complex representation the algebra $\mathbb{R}_{3,0}$
generates by multiplication the elements
$j\sigma_k$, $i\sigma_k$, and $ij$. 
The algebra can be complexified with either the hyperbolic or the complex unit,
which provides the additional elements $i$, $j$, $\sigma_k$, and $ij\sigma_k$.
The full structure is equivalent to the complex Clifford
algebra $\bar{\mathbb{C}}_{3,0}$. The sixteen element complexified Pauli algebra has been used in
considerations on relativistic quantum physics by Edmonds \cite{Edm72}.

\section{Hyperbolic complex spaces and groups}
\label{spaces}
In Sec.~\ref{alg} the hyperbolic unit $j$ has been introduced. This
implies that the complex numbers can be extended to the hyperbolic
complex numbers $z\in\mathbb{H}$
\be
\label{beg}
z=x+iy+jv+ijw\;,\hspace{0.5cm}x,y,v,w \in\mathbb{R}\;.
\ee
This commutative ring has been applied earlier by Hucks \cite{Huc93} to the Dirac formalism
and the representation of Dirac spinors.

It is thus possible to construct vector spaces of hyperbolic complex
numbers and groups, which act on these vector spaces. 
The scalar product of elements $\varphi,\psi\in \mathbb{H}^n$ is defined with 
the conjugation anti-involution, represented as transposition
and change of sign of hyperbolic and complex units
\be
(\varphi,\psi)\rightarrowtail\bar{\varphi}\psi\in\mathbb{H}\;.
\ee
In coordinate representation this is written as $\bar{\varphi}\psi=\bar{\varphi}_i\psi^i$.
The vectors $\varphi$ and $\psi$ can be considered as spinors with
an element of the spin group acting on the 
spinors according to
\be
g:\mathbb{H}^n\rightarrow \mathbb{H}^n;\;\psi\rightarrowtail \bar{g}\psi,\;\psi\in \mathbb{H}^n\;.
\ee
Matrix and vector indices have been omitted.
Under this transformation, the scalar product of two spinors changes into
\be
\label{trafo2}
(\varphi,\psi)\rightarrowtail (\bar{g}\varphi,\bar{g}\psi)= \bar{\varphi}g\bar{g}\psi\;.
\ee
The scalar product remains invariant for elements of the
hyperbolic unitary group
\be
U(n,\mathbb{H})=\left\{g\in \mathbb{H}(n)) : g\bar{g}=1\right\}\;,
\ee
where $\mathbb{H}(n)$ denotes a hyperbolic complex $n\times n$ matrix.
As usual, the special unitary restriction is defined as the subgroup
which includes the elements with unit determinant
\be
SU(n,\mathbb{H})=\left\{g\in U(n,\mathbb{H})) : det(g)=1\right\}\;.
\ee

The elements of the group transformations $g$ may be represented 
for both groups in the form
\be
\label{gelement}
g=\exp{(-i\phi + j\xi)}\;,
\ee
where the transformation parameters can be expanded in terms
of the Lie algebra $\mathfrak{g}$
\be
\label{gener}
\phi= \phi^i\tau_i,\;\;\;\xi= \xi^i\kappa_i,\;\;\;\;\;\;\tau_i,\kappa_i\in\mathfrak{g}\;.
\ee
In the case of $SU(n,\mathbb{H})$ the Lie algebra is represented by
traceless matrices, whereas for $U(n,\mathbb{H})$ the Lie algebra
also includes the unit matrix. Based on the sign conventions of Eq.~(\ref{gelement})
there is the  relationship
\be
\label{gener2}
\kappa_i=ij\tau_i\;.
\ee
This indicates that the Lie algebras of $U(n,\mathbb{H})$ and $SU(n,\mathbb{H})$
are complex extensions of the Lie algebras of $U(n,\mathbb{C})$ and $SU(n,\mathbb{C})$.
In this case, the complex extension is done with $ij$. If one
uses the complex unit $i$ alone, the complex extension is leading to the
groups $GL(n,\mathbb{C})$ and $SL(n,\mathbb{C})$. The hyperbolic unitary
groups are therefore isomorphic to the complex linear groups,
which has been pointed out earlier by Zhong \cite{Zho84,Zho85,Zho92}.
One may be irritated, why a unitary group can be isomorphic to a non-unitary linear group.
The answer is that the linear groups are, in fact, also unitary \cite{Por95}.

Due to this isomorphism, the structure of the hyperbolic unitary groups is well understood.
The Lie algebra of $SU(n+1,\mathbb{H})$ has rank $n$ and dimension $n(n+2)$.
The corresponding Dynkin diagram is $A_n$. The
group representations are derived as usual. 
For concrete calculations, one may adopt
the knowledge from common textbooks on group theory. One only has to keep
in mind that in certain locations the complex unit $i$ must be replaced 
by $ij$. 

As an example, the finite dimensional representation $\vert\sigma \,\rho\rangle\equiv\vert (s\sigma) ,(r\rho)\rangle$ 
of the relativistic spin group $SU(2,\mathbb{H})$, where $(s\sigma)$ and $(r\rho)$ indicate
the quantum numbers of $SU(2,\mathbb{C})\times SU(2,\mathbb{C})$,
gives rise to the relations
\bea
\label{tprop}
J_3\vert \sigma \,\rho\rangle &=& (\rho+\sigma)\vert \sigma \,\rho\rangle\;,\nonumber\\
K_3\vert \sigma \,\rho\rangle &=& ij(\rho-\sigma)\vert \sigma \,\rho\rangle\;.
\eea
$J_3$ and $K_3$ are the third components of the generators of the relativistic spin group \cite{Tun85}.

\section{Wave equations}
One may extend the considerations of Secs.~\ref{alg} and \ref{spaces} to 
vector and spinor fields in relativistic space-time.
The scalar product for vector fields is generalized to
\be
(A,B):= \int\frac{1}{2}tr(\bar{A}B)\;.
\ee
The integration covers relativistic space-time.
The vector fields map every point in space-time
to an element of the relativistic paravector algebra. One may introduce
the notation $A,B\in {}^{\ast}W$ to indicate this \cite{Goc87}.
For spinor fields, the definition
\be
(\varphi,\psi):= \int\bar{\varphi}\psi
\ee
is introduced. The action for a massless gauge field and a massive spinor field, 
which are not interacting with each other at this stage, can then be defined as
\be
\label{action}
S[A,\psi]=(A,M^2A)+(\psi,(M^2-m^2)\psi)\;,
\ee
where $m^2\in\mathbb{R}$ denotes the squared particle mass. 

The Lagrangian for free
noninteracting spinor fields can be derived from the above definitions as
\be
\label{lagrange}
\mathcal{L}=\bar{\psi}(M^2-m^2)\psi\;.
\ee
The mass operator is expressed in terms of the relativistic paravector algebra.
Equation~(\ref{trafo2}) indicates how
an operator between two spinor functions needs to be formulated in order to provide
an invariant expression. The mass operator is thus defined as
\be
M^2=p\bar{p}\;.
\ee
The components of the momentum operator are given as $p=p^\mu e_\mu=i\partial^\mu e_\mu$,
one finds $\bar{p}^\mu=p^\mu$.

The equations of motion can be derived in the usual way.
For spinor fields, this results in
\be
\label{mass}
M^2\psi=m^2\psi\;.
\ee
Without interactions, this is just the Klein-Gordon equation
for spinor fields.
The equation of motion for the gauge field $A\in {}^\ast W$ is given by
\be
\label{homog}
M^2A=0\;.
\ee
In order to make the relationship to hyperbolic complex mathematics more explicit,
the basis vectors $e_\mu$ included in the mass operator need to be evaluated
according to Eq.~(\ref{basis}). A short calculation transforms Eq.~(\ref{homog}) into the
homogenous Maxwell equations in the hyperbolic complex form \cite{Ulr05Fund}.

Rodrigues and Capelas de Oliveira used the notion of a Clifford bundle \cite{Rod07}.
In this sense the gauge field $A$ and the spinor field $\psi$
can be understood as sections of the Clifford bundle and the
Spin-Clifford bundle of
$\mathbb{R}_{3,0}$. 

One may ask why another abstract formalism is invoked to represent
wave equations. The geometric structure of the Maxwell equations is well
understood in the framework of differential geometry. Why is so much emphasis
put on the mass operator? The reason is that the 
mass operator naturally arises as the Casimir operator
of the Poincar\'e group, the group of rotations, and
translations in relativistic space-time. The only ingredients into
the theory are thus a metric vector space and its
isometries as the corresponding set of
transformations. The basic wave equation then emerges in a natural way
from the geometric properties of the metric space, which is not so obvious in the
conventional formalisms.

Consider the irreducible representations of the Poincar\'e group
$\vert p\rangle\vert \sigma \,\rho\rangle$, where $\vert p\rangle$ is the
eigenket of the momentum operator and $\vert \sigma \,\rho\rangle$ is the finite
dimensional representation of the Lorentz group introduced in Eq.~(\ref{tprop}).
The fundamental $(1/2,0)\oplus (0,1/2)$ representation, which can be represented 
according to Hucks \cite{Huc93} as hyperbolic complex two component spinors,
implies a $2\times 2$ matrix framework
of operators acting on these objects. 
The mass operator defined in Eq.~(\ref{mass}) is therefore the natural representation of the
Casimir operator related to $\vert p\rangle$ within the spin structure given by this Lorentz representation.
The hyperbolic Pauli algebra, which realizes this representation, thus takes into account the basic
spin structure induced by relativistic space-time as the underlying metric space and the Poincar\'e
transformations as the underlying set of transformations.

\section{Hyperbolic complex Yang-Mills fields}
\label{yang}
Interactions can be introduced into the mathematical framework
in the usual way. The covariant derivative is defined as
\be
\label{subs}
p\rightarrow p^\prime=p+A\;,
\ee
which ensures the invariance under gauge transformations.
A coupling constant has been omitted for simplicity.
The action and the Lagrangian in  Eqs.~(\ref{action}) and (\ref{lagrange})
remain formally invariant. However, the mass operator for the spinor
field contributions is modified by the minimal substitution
\be
M^2=(p+A)(\overline{p + A})\;.
\ee
This mass operator corresponds to the quadratic Dirac operator with interactions.
Again the basis vectors of the mass operator need to be evaluated to show the hyperbolic complex structure of the equation \cite{Ulr05Fund}.
The mass operator for the gauge bosons remains unchanged.
However, the equations of motion are changed to
\be
M^2A=-J\;,
\ee
with the current $J$ as the source term of the interacting
field. 

The fields can be generalized to Yang-Mills fields. One may extend the gauge
transformations even further to include hyperbolic complex unitary
transformations as in  Eq.~(\ref{gelement}). With the parameter
definitions given in Eq.~(\ref{gener}), the gauge fields then obtain the form
\be
A=G+H\;,
\ee
including the Yang-Mills type of structure
\be
G=G^i \tau_i,\hspace{0.5cm} H=H^i \kappa_i\;.
\ee
As in Eq.~(\ref{gener2}), the generators $\kappa_i$ are hyperbolic 
complex extensions of the generators $\tau_i$ by the factor $ij$.

One may ask now whether this is only a mathematical construction
or whether there are indications for new type
of interactions, which go beyond the interactions
obtained by the usual complex gauge groups.

\section{New interactions}
The hyperbolic complex gauge transformations introduced in Sec.~\ref{yang} give rise for the hypothesis of an extended set
of interactions. Instead of interactions based on complex gauge groups,
one may think of hyperbolic complex principal bundles.
For the hyperbolic complex group $U(1,\mathbb{H})$ it has been proposed in Ref.~\cite{Ulr06Gravi}
to understand a Maxwell-like theory of gravitation as the hyperbolic
complex extension of electromagnetism. This idea
is based on a model of Majern\'ik \cite{Maj71}, who extended current and fields
of the Maxwell theory to be proportional to the complex unit $i$.
In Ref.~\cite{Ulr06Gravi} this idea is modified to extend the Maxwell theory
by the factor $ij$. Comparison with the experiment is given by the analysis of 
Singh \cite{Sin82}, who explained the precession of the perihelion of a planet,
the deflection of light in the
gravitational field of a star, and the gravitational redshift based on
velocity dependent spatial components of the relativistic vector potential.
The spatial components correspond to a boosted potential, with the Newton potential introduced in the rest system 
as the zero component of a relativistic vector potential.

For the case $SU(2, \mathbb{H})$, one might think on the first sight that the new
$1 + ij$ form of the combined complex and hyperbolic gauge interaction
gives a natural representation for the symmetry breaking structure of the weak interaction.
Note that the hyperbolic unit $j$ can be identified with the  $\gamma_5$ Dirac operator \cite{Huc93,Ulr05Fund}.
However, projectors to left- and right-handed 
spinors in the hyperbolic complex approach are proportional to $1\pm j$.
Therefore, this slight but significant difference excludes
this relationship.

A hyperbolic complex counterpart of quantum chromodynamics (QCD) may be assessed by intuitive
arguments. If one assumes seriously a relationship between gravitation and electromagnetism as above,
the hyperbolic complex counterpart of QCD should be much weaker than QCD, it should act
on a longer length scale because the electromagnetic forces are more or less confined to neutral objects such as planets, and it should be attractive for composite objects,  which are uncharged. These relationships hold between
electromagnetism and gravitation and may be transferred to the high energy scale.
In particle and nuclear physics these requirements are satisfied
by quantum hadrodynamics (QHD), the theory which is built on hadrons as
effective degrees of freedom. One may therefore ask whether QHD and QCD
stand in a similar relationship like gravitation, as a 
hyperbolic complex Maxwell-like theory,
and quantum electrodynamics. 
However, the mathematical structure of QHD and QCD is very different.
In addition, QHD is an effective theory, whereas QCD and gravitation are mostly considered
as fundamental. Therefore, it seems to be inappropriate to place QHD in the same scheme as other fundamental interactions.
Nevertheless, it should be mentioned that Zee and Adler \cite{Zee79, Adl80, Adl82} considered gravitation as an effective theory,
and also QCD may be considered as effective in the light of the preon models.

This discussion shows that the straightforward extension of
gauge theories to hyperbolic complex transformations leaves many questions open and is possibly not of
any physical relevance. One should keep in mind the noncompactness of the group structure,
which might exclude these transformations in this context.
As mentioned earlier, the straightforward hyperbolic complex generalization 
of the orbital angular momentum presented in Ref.~\cite{Ulr05Fund} does
not lead to physical relevant functions because these functions
are not solutions of the Klein-Gordon equation in the corresponding
representation. In this sense, it should be the goal of future investigations
to understand exactly whether hyperbolic complex contributions
have physical relevance for the internal symmetries of a particle state,
and if not to understand exactly why.

\section{Summary}
The hyperbolic complex approach to relativistic quantum physics 
presented in earlier publications has been discussed and summarized.
The main benefit is the improved representation of the spinor structure
and the possibility to consider the basic wave equation
as the spinor representation of a Casimir operator
und thus emphasize the foundation in the geometric structure of space-time and the Poincar\'e group.

Attempts have been made to look for new physics related to
hyperbolic complex extensions of the gauge groups. Though a simple model
can be given to understand a Maxwell-like theory of gravitation as the hyperbolic complex extension of electromagnetism,
it is not straightforward to apply this model with physical relevance to interactions of Yang-Mills type.
Further investigations are required to clarify this possibility.

\end{document}